# PERFORMANCE ANALYSIS OF RESOURCE SCHEDULING IN LTE FEMTOCELLS NETWORKS

Samia Dardouri[1] and Ridha Bouallegue[2]

[1]National Engineering School of Tunis, University of Tunis El Manar, Tunisia
samia_telecom@yahoo.fr
[2] Innov'COM Laboratory, Sup'Com, University of Carthage, Tunisia
ridha.bouallegue@ieee.org

## ABSTRACT

*3GPP has introduced LTE Femtocells to manipulate the traffic for indoor users and to minimize the charge on the Macro cells. A key mechanism in the LTE traffic handling is the packet scheduler which is in charge of allocating resources to active flows in both the frequency and time dimension. So several scheduling algorithms need to be analyzed for femtocells networks. In this paper we introduce a performance analysis of three distinct scheduling algorithms of mixed type of traffic flows in LTE femtocells networks. The particularly study is evaluated in terms of throughput, packet loss ratio, fairness index and spectral efficiency.*

## 1. INTRODUCTION

Recently demands of users for wireless data communications in cellular networks are rapidly increasing as fascinating mobile devices and mobile applications. One of the important challenges for LTE systems is to ameliorate the indoor coverage and enhance high-data-rate services to the users in a performant way and at the same time to improve network capacity [1].

However, The LTE femtocells are referred to as Home evolved Node Bs (HeNBs) which is a femtocell base station and extensive evolved UMTS terrestrial radio access network (E-UTRAN) architecture are defined to support femtocells in the LTE system which is illustrated in Fig. 1 [2]. HeNB contains most functionalities of eNB and is linked to cellular core networks through the existing Internet access and HeNB gateway. The femtocell is a home base station which is meant to be deployed in homes or companies to rice the mobile network capacity or offer mobile network coverage where none exists [3],[4].

Therefore many searchers discuss reuse of wireless resources in macro and femtocell in order to enhance utilization of the limited wireless resources [2]. Radio resources in both time and frequency domains can be individually allowed to macro and femtocells. All of the scheduling algorithms are developed by considering the dynamics of the macrocell wireless networks.

However femtocell networks are different from macrocell networks in terms of network coverage range number of simultaneous users served total transmission power and total available networks resources [5]. So several scheduling methods need to be developed for femtocell networks and before that the performance of suggested schedulers need to be analyzed such as those presented in [6], [7] and [8].

Furthermore resource allocation can be classified in the light of applications demands like that the performance of scheduling algorithms extremely relies on the type of mixed flows [9], [10]. In order to implement higher performance of system, it is important to select appropriate algorithms depending on flows applications. The flows can be mixture of Real-Time as well as

Non-Real-Time flows due to the evaluation of basic services of LTE, including voice service, data service, and live video service [11], [12], [13].

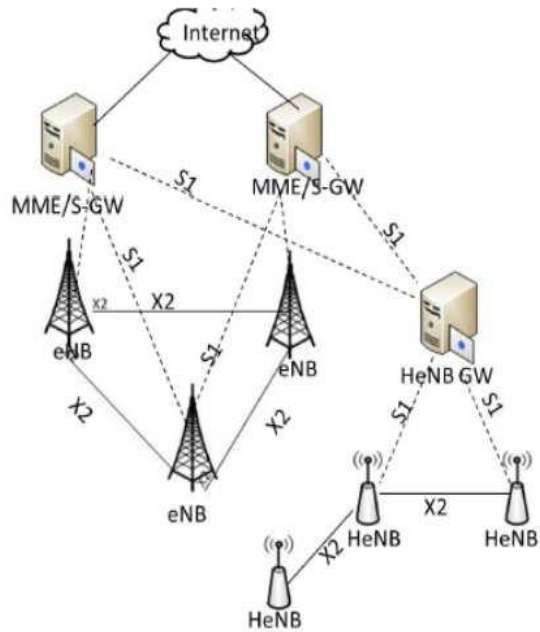

Fig. 1: Architecture of LTE Femtocell Networks.

In this work, we study several scheduling algorithms of different types of downlink traffic in LTE system. we apply a metrics which enables fast evaluation of performance metrics such as mean flow transfer times manifesting the effect of resource allocation techniques using LTE-Sim simulator [14]. This paper examines the radio resource scheduling in LTE femtocell wireless network and is organized as follows: Section II represents the radio resource allocation in LTE and presents scheduling algorithms with mathematical formula. Then Section III exposes the simulation results and performance analysis. Section IV is consecrated to the conclusion.

## 2. RADIO RESOURCES SCHEDULING

In an LTE system, the spectrum is separated into fixed sized chunks called Resource Blocks (RBs). One or more RBs can be affected to service an application request subject to the availability of the resource and network policies. Unless stated otherwise, the ensuing discussion is based on the assumption that the underlying system under consideration is LTE [15].

The radio resource scheduling in LTE is assigned in both time and frequency domain. The LTE air interface elements are given in Fig. 2. In time-domain the DL channels in air interface are separated into Frames of 10 ms each. Frame includes 10 Sub frames each of 1 ms. Each subframe interval is attributed to as Transmission Time Interval (TTI). Each subframe consists of 2 Slots of 0.5ms. In frequency domain the total available system bandwidth is separated into sub-channels of 180 kHz with each sub-channel including 12 successive equally spaced OFDM sub-carriers of 15 KHz each [16].

A time-frequency radio resource covering over 0.5 ms slots in the time domain and over 180KHz sub-channel in the frequency domain is named Resource Block (RB). The LTE effectuate in the bandwidth of 1.4 MHz up to 20 MHz with number of RBs ranging from 6 to 100 for bandwidths 1.4 MHz to 20 MHz respectively [2].

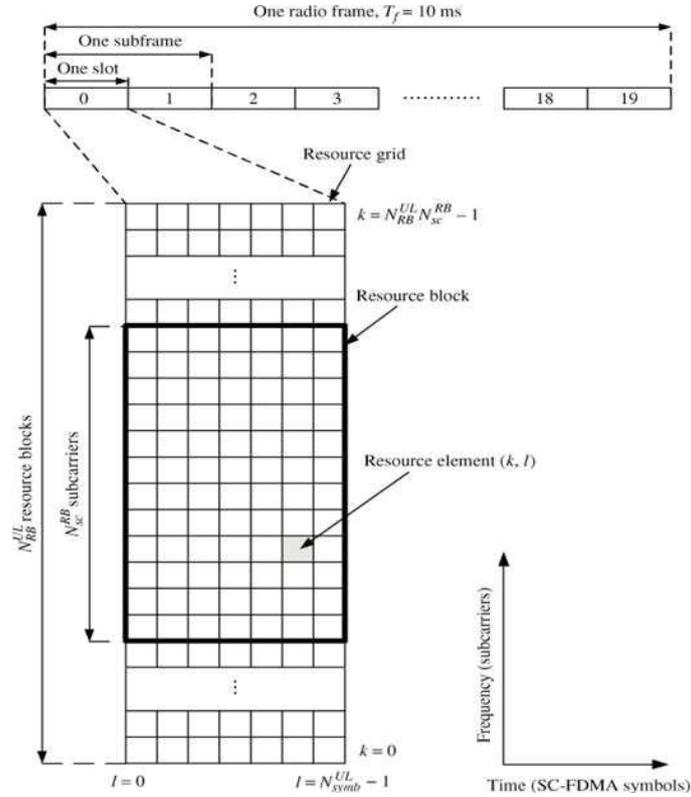

Fig. 2: The structure of the downlink resource grid.

In the following, the description of different scheduling algorithms were used in all simulation scenarios, these are: PF as well as Log-Rule and FLS.

## 2.1. Proportional Fair (PF)

The PF scheduling algorithm [16] provides a good trade-off between system throughput and fairness by selecting the user. For this algorithm, the metric is given by:

$$j = \max_i \frac{u_i(t)}{\bar{u}_i} \quad (1)$$

Where $\bar{u}_i$ is the rate corresponding to the mean fading level of user $i$ and $u_i(t)$: is the state of the channel of user i at time t.

## 2.2. Logarithmic rule (LOG rule)

The log rule has been presented in [7]. The log rule is explained as follows:

$$j = \max_i \log(1 + a_i q_i) \frac{u_i(t)}{\bar{u}_i} \quad (2)$$

Where $u_i(t)$ and $\bar{u}_i$ are the same parameters already presented in PF scheduler. The value of qi represents the length queue. $a_i = 5/d_i$ Where $d_i$ is the maximal delay target of the i-th user's flows.

### 2.3. Frame Level Scheduler (FLS)

The FLS [8] is a two-level scheduling techniques with one upper level and lower level. Two different algorithms are implemented in these two levels. At the upper level, a discrete time linear control law is used every LTE frame (i.e.10 ms). It computes the total amount of data that real-time flows should transmit in the following frame in order to satisfy their delay constraints. When FLS finishes its task, the lowest layer scheduler works every TTI to assign resource scheduling using the PF schemas. PF considers the bandwidth requirements proposed by the FLS. Firstly, the lowest layer scheduler allocates RBs to those UEs that experience the best CQI (Channel Quality Indicator) and then the rest ones are considered. If any RBs left unattached, they would be allocated to best-effort flows.

## 3. SIMULATION RESULTS AND PERFORMANCE ANALYSIS

This section describes the simulations results to evaluate three schedulers performance and discusses the results obtained from experiments conducted using the LTE simulator developed in [14]. Before the discussion on the simulation results, the simulation scenario and the performance metrics are given. Then,We compare three previously proposed schedulers in terms of their achieved throughput, fairness index, Packet Loss Ratio and spectral efficiency in different number of users.

### 3.1. Simulation Scenario

In these simulations we designed a scenario including one macrocell and 56 buildings situated as in a urban scenario. The cell itself has one enodeB which transmits using an omnidirectional antenna in a 5 MHz bandwidth. Each UE uses at same time a video flow a VoIP flow and a best effort flow as shown in Figure 3. The video flow are based on realistic video trace files with a rate of 128 kbps was used. For VoIP a G.729 voice stream with a rate of 8.4 kbps was considered. The LTE propagation loss model is organized by four different models (shadowing, multipath, penetration loss and path loss):

• Path loss: PL = 128.1 + 37.6 log10(d) where d is the distance between the UE and the eNodeB in km.

• Multipath: Jakes model.

• Penetration loss: 10 dB.

• Shadowing: log-normal distribution, with mean 0 dB and standard deviation of 8 dB.

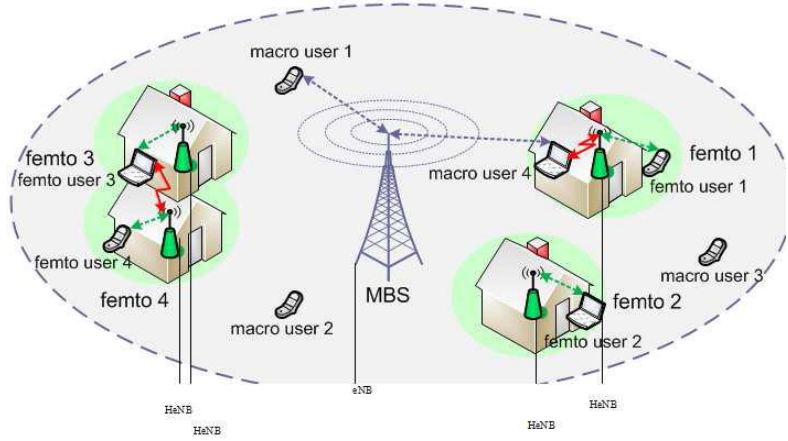

Fig. 3: LTE simulated scenario with video, VoIP and best effort flows.

### 3.2 Performance Metrics

When evaluating the Quality of service, several metrics are used in this work such as Average Throughput, packet loss Ratio, fairness index and spectral efficiency. Three metrics are used in this paper and are explained as follows:

• Average Throughput per user. This metric presents the average rate of successful message delivery over physical channel. It is elaborated by dividing the size of a transmitted packet by the time it selects to transfer the packets per each user. We chose this metric to examine the impact of throughput when the number users increase.
• Packet Loss Ratio (PLR). This metric try to measure the percentage of packets of data transmitting across a physical channel which fail to reach their destination. Also there exist packet losses caused by buffer overflows [19].
• Fairness Index. In order to acquire an index related to the fairness level we use the Jains fairness index method [20].

$$FI = \frac{(\sum_1^N x_i)^2}{N \sum_1^N x_i^2} \qquad (3)$$

Where $x_i$ is the throughput assigned to user i among N competing flows.

### 3.3 Simulation Results

Now, we examine the effect of the femtocells unfolding in urban environments. For this, a typical urban Scenario without femtocells and urban Scenario with femtocells exposed are compared. From Fig.4. It is possible to observe that the PLR of best effort flows increase of all sheduling algorithms in case of adoption of femtocells. With reference to VoIP flows, we examined that they achieved smaller PLR without femtocells.

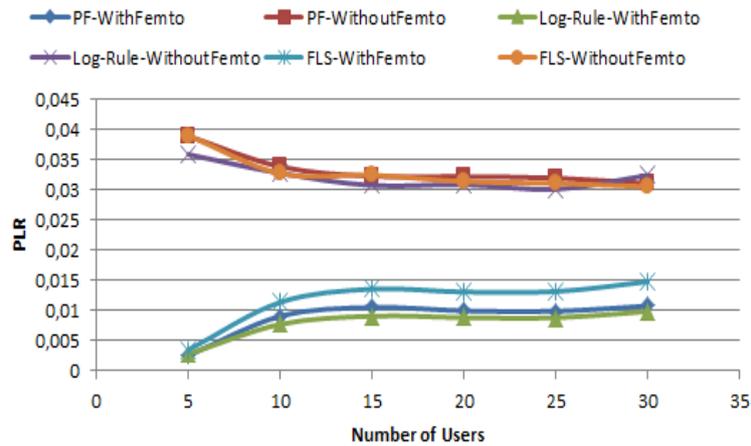

Fig. 4. Packet Loss Ratio of best effort flows

The VOIP flows experience has lower PLRs than Video flows are illustrated in Fig. 5. Initially, Fig. 6 shows that the PLR increased as the number of UEs in the cell increased, which is obviously due to the increased load on the network. The FLS scheduler is outperformed than LOG-Rule and FLS algorithms in terms of PLR, especially for video streaming. FLS has achieved the lowest PLR in video and VoIP flows.

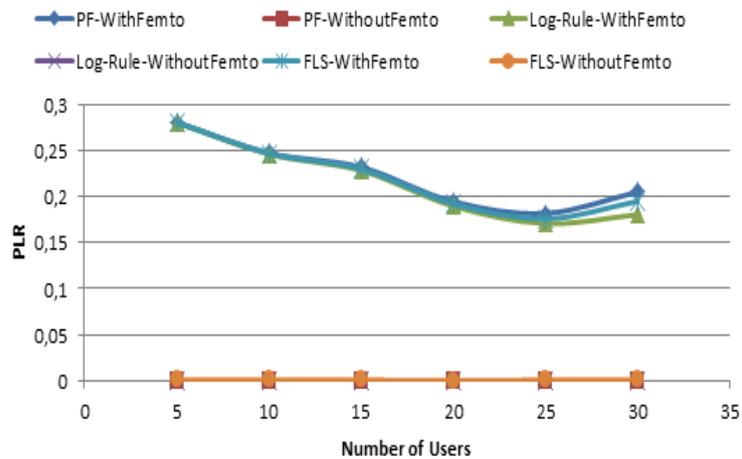

Fig. 5. Packet Loss Ratio of VoIP flows

Then we analysed the throughput obtained by different flows for the presented scheduling algorithms. In order to do that, we analysed the overall system throughput for each scenario considering the number of UEs in both the macrocells and femtocells scenario. In all the cases, it is figured that the PF Algorithm shows good throughput performance for best effort flows. This result is shown in Fig. 7.

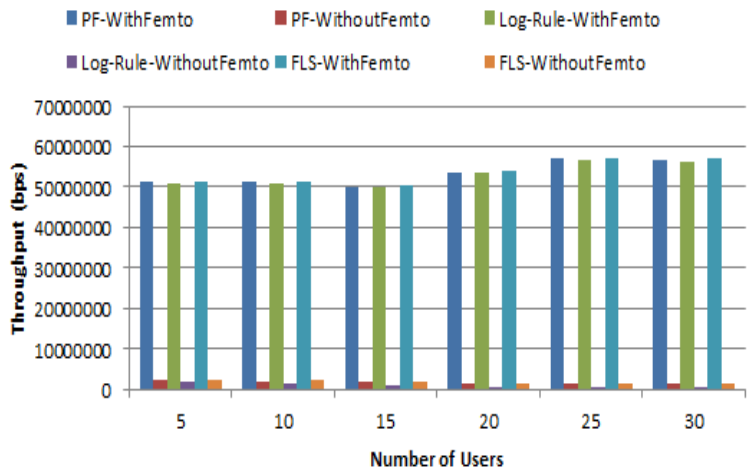

Fig. 6.  Throughput of best effort flows.

Throughput, can also analysed by evaluating the performance of real time flows. The Fig.8 shows the throughput obtained by VoIP flow versus the number of UEs in the cell. We see that when number of users in the cell increases, the Log-Rule and FLS schedulers maintain a high throughput compared with the PF. Finally, it is important to remark that VoIP flows experience exactly smaller throughput than video ones is shown in Fig. 9. The performance of FLS scheduler is the greater. In such case we need to choose an algorithm having comparatively higher throughput especially in Real time flows.

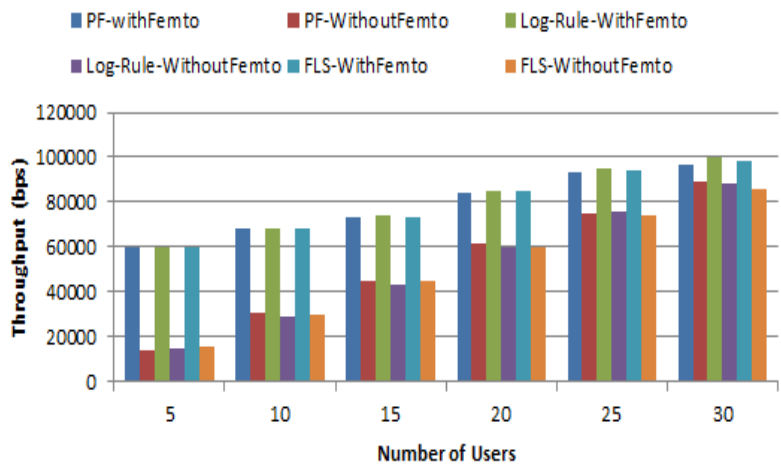

Fig. 7.  Throughput of VoIP flows.

Resource scheduling techniques should optimally scale between fairness in order to ensure QOS. Although, Table 1 demonstrate that the FLS scheduling discipline can maintain a high level of fairness index in both femtocells and macrocells scenario. Table 2 presents the fairness index experienced by VOIP flows. It shows that the FLS scheduler degree of fairness performance is higher compared to PF and LOG-Rule scheduler in both femtocells and macrocells scenario.

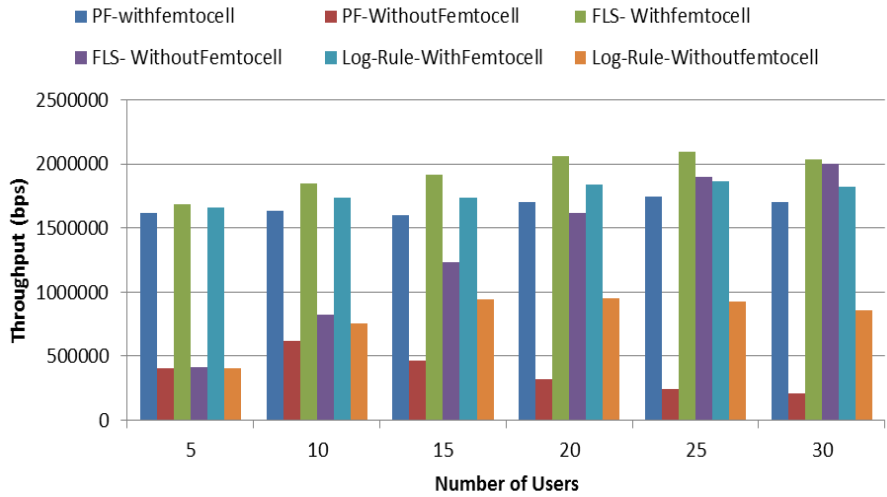

Fig. 8. Throughput of video flows.

Table 1: FAIRNESS INDEX VALUE OF VIDEO FLOWS.

| Scenario | Without Femtocell | | | With Femtocell | | |
|---|---|---|---|---|---|---|
| Number of users | PF | FLS | LOG-Rule | PF | FLS | LOG-Rule |
| 5 | 0.28 | 0.35 | 0.34 | 0.39 | 0.4 | 0.39 |
| 10 | 0.10 | 0.26 | 0.19 | 0.26 | 0.3 | 0.29 |
| 15 | 0.10 | 0.26 | 0.17 | 0.1461 | 0.33 | 0.32 |
| 20 | 0.07 | 0.24 | 0.15 | 0.12 | 0.34 | 0.30 |
| 25 | 0.06 | 0.19 | 0.12 | 0.10 | 0.31 | 0.25 |
| 30 | 0.05 | 0.16 | 0.09 | 0.08 | 0.32 | 0.23 |

Table 2: FAIRNESS INDEX VALUE FOR VOIP FLOWS.

| Scenario | Without Femtocell | | | With Femtocell | | |
|---|---|---|---|---|---|---|
| Number of users | PF | FLS | LOG-Rule | PF | FLS | LOG-Rule |
| 5 | 0.33 | 0.33 | 0.33 | 0.37 | 0.38 | 0.39 |
| 10 | 0.34 | 0.35 | 0.35 | 0.39 | 0.39 | 0.39 |
| 15 | 0.28 | 0.28 | 0.28 | 0.32 | 0.31 | 0.32 |
| 20 | 0.29 | 0.30 | 0.30 | 0.33 | 0.34 | 0.34 |
| 25 | 0.30 | 0.32 | 0.31 | 0.34 | 0.34 | 0.34 |
| 30 | 0.27 | 0.30 | 0.29 | 0.32 | 0.32 | 0.32 |

Finally, Table 3 shows that the FLS scheduling algorithm can maintain a high level of fairness index especially in scenario with Femtocells.

Table 3: FAIRNESS INDEX VALUE OF BEST EFFORT FLOWS.

| Scenario | Without Femtocell | | | With Femtocell | | |
|---|---|---|---|---|---|---|
| Number of users | PF | FLS | LOG-Rule | PF | FLS | LOG-Rule |
| 5 | 0.2 | 0.2 | 0.2 | 0.2 | 0.2 | 0.2 |
| 10 | 0.18 | 0.17 | 0.18 | 0.25 | 0.24 | 0.25 |
| 15 | 0.18 | 0.18 | 0.18 | 0.26 | 0.25 | 0.26 |
| 20 | 0.16 | 0.16 | 0.16 | 0.23 | 0.20 | 0.23 |
| 25 | 0.17 | 0.17 | 0.17 | 0.25 | 0.22 | 0.25 |
| 30 | 0.14 | 0.14 | 0.14 | 0.24 | 0.19 | 0.24 |

Finally, Fig.10 shows that the total cell spectral efficiency increases as long as the number of users increases. It can notice that the use of femtocells increases spectral efficiency in LTE systems.

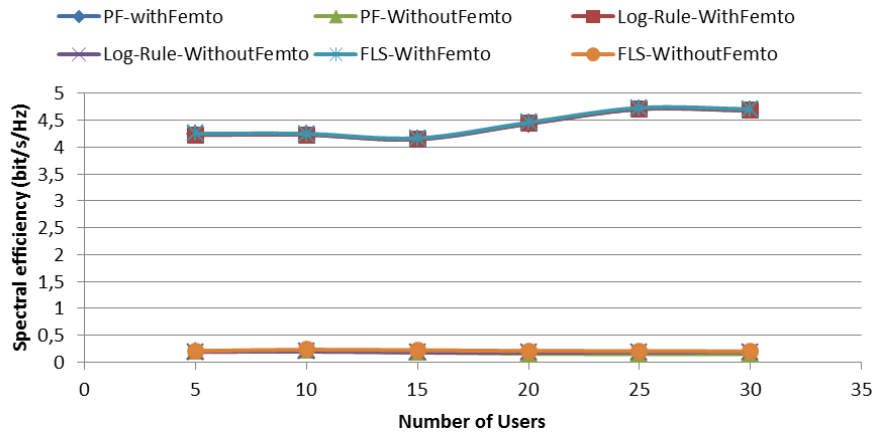

Fig. 9: Spectral efficiency

## 3. CONCLUSIONS

In this paper, we studied the resource allocation problem in LTE femtocells networks. Using the LTE-SIM simulator, the study compares the performance of scheduling schemas such as average system throughput, PLR, and fairness via video and VoIP traffic in femtocells scenario. Our simulation shows FLS performs better than both PF and Log-Rule with respect to throughput satisfaction rate. For RT Traffic, PF shows the highest PLR value, the lowest attained throughput and a high delay when the cell is charged; thus this algorithm could be suitable for non-real-time flows but is inappropriate to manipulate real time multimedia services.

We found that FLS always reaches the lowest PLR in all used scenarios, among all those schemas that try to ensure attached delay but at the cost of reducing resources for best effort flows. Adoption of femtocells can increase the Overall system throughput.

Upcoming work will examine also the more challenging problem of scheduling such as the uplink direction using multicells scenario.

ACKNOWLEDGEMENTS

This work was sustained in part by Laboratory InnovCOM of Higher School of Telecommunication of Tunis, and National Engineering School of Tunis.

**Authors**

**DARDOURI Samia**
Received the B.S. degree in 2009 from National Engineering School of Gabes, Tunisia, and M.S. degree in 2012 from National Engineering School of Tunis. Currently he is a Ph.D student at the School of Engineering of Tunis. He is a researcher associate with Laboratory at Higher School of Communications (SupCom), University of Carthage, Tunisia. Her Research interests focus on scheduling algorithms and radio resource allocation in LTE systems.

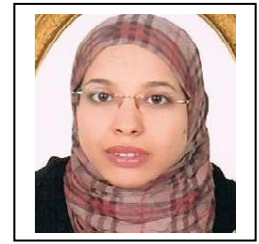

**PR. RIDHA BOUALLEGUE**

Received the Ph.D degrees in electronic engineering from the National Engineering School of Tunis. In Mars 2003, he received the Hd.R degrees in multiuser detection in wireless communications. From September 1990 He was a graduate Professor in the higher school of communications of Tunis (SUP'COM), he has taught courses in communications and electronics. From 2005 to 2008, he was the Director of the National engineering school of Sousse. In 2006, he was a member of the national committee of science technology. Since 2005, he was the laboratory research in telecommunication Director's at SUP'COM. From 2005, he served as a member of the scientific committee of validation of thesis and Hd.R in the higher engineering school of Tunis. His recent research interests focus on mobile and wireless communications, OFDM, OFDMA, Long Term Evolution (LTE) Systems. He's interested also in spacetime processing for wireless systems and CDMA systems.

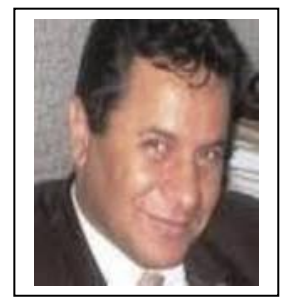